\RequirePackage{fix-cm}
\documentclass[smallextended]{svjour3}       
\smartqed  
\usepackage{graphicx}
\begin{document}

\title{Vibration can enhance stick-slip behavior for granular friction
}


\author{Abram H. Clark, Robert P. Behringer, and Jacqueline Krim   
}


\institute{A. H. Clark \at
              Naval Postgraduate School \\
              Department of Physics \\
              Monterey, CA 93943 \\
              \email{abe.clark@nps.edu}    \\        
         \and
 			R. P. Behringer \at
              Duke University \\
              Department of Physics \\
              Durham, NC 27708 \\
 	\and
			J. Krim \at
			North Carolina State University \\
			Department of Physics \\
			Raleigh, NC 27695 \\
			\email{jkrim@ncsu.edu}
}

\date{Received: date / Accepted: date}

\maketitle

\begin{abstract}
We experimentally study the frictional behavior of a two-dimensional slider pulled slowly over a 
granular substrate comprised of photoelastic disks. The slider is vibrated at frequencies ranging from 
0 to 30 Hz in a direction parallel to sliding. The applied vibrations have constant peak acceleration, which results in constant average friction levels. Surprisingly, we find that stick-slip behavior, where stress slowly builds up and is released in intermittent slips, is enhanced as the frequency of vibration is increased. Our results suggest that increasing the frequency of vibration may help to combine many smaller rearrangements into fewer, but larger, avalanche-like slips, a mechanism unique to granular systems. We also examine the manner in which the self-affine character of the force curves evolves with frequency, and we find additional support for this interpretation.
\end{abstract}

\section{Introduction}
\label{intro}

The frictional response of granular material has broad relevance in earth science (\textit{e.g.}, fault mechanics~\cite{scott1994apparent,daniels2008force,hayman2011granular}, landslides~\cite{darve2000instabilities}, sediment erosion~\cite{clark2017,pahtz2018}) and industry (\textit{e.g.}, food~\cite{peleg1977flowability,teunou1999effect}, pharmaceuticals~\cite{ketterhagen2009process}, and detergents~\cite{tardos1997}). Moreover, yield and flow of granular materials in geologic or industrial settings is often accompanied by latent vibrations in the system~\cite{elst2012,taylor2017}. The interplay between vibrations and granular friction has important implications for dynamic earthquake triggering~\cite{johnson2005nonlinear,brodsky2014uses} and other avalanche-like behavior~\cite{dahmen2011}. 

Friction in granular systems is not a simple material property: for example, the material friction coefficient $\mu$ is only weakly dependent on the grain-grain friction coefficient~\cite{dacruz2005,kamrin2014}. Instead, an effective friction coefficient for granular materials arises from the ability of the grains to form anisotropic force networks~\cite{majmudar2005,peyneau08}, often called force chains. These structures can be correlated over large distances and can thus slip and flow in collective and complex ways that are difficult to predict~\cite{nichol2010,dahmen2011,henann2013,bouzid2013,clark2018}. Thus, near yield, the effect of vibration is magnified, as marginally stable force chains can be disrupted and reorganize into either weaker or stronger configurations. Vibration can lead to compaction, dilation, and/or fluidization of the granular material, depending on the magnitude of the acceleration that the grains experience.~\cite{dijksman2011,Kirsch2016}. Factors such as the peak velocity experienced by the vibrating grains and the direction of vibrations also affect the dynamics, and the relative contribution of each is a matter of great current interest~\cite{Lastakowski2015,Popov2010}.



Previous studies~\cite{braiman1999,Popov2010,dijksman2011,krim2011stick,Lastakowski2015,zadeh2017,zadeh2018crackling} have shown that increasing amplitude of vibration and increasing slider speed can each result in reduced stick-slip behavior and average granular friction levels.
In this study, we demonstrate that, under certain conditions, increased frequency of vibration can surprisingly enhance stick-slip behavior. We present results for a slider pulled at constant speed of 5~mm/s over a granular bed. The slider is vibrated in a direction parallel to its motion by means of an electromagnetic shaker. We vary $f$ between 0~Hz and 30~Hz and set the displacement amplitude $A$ of the shaking such that the dimensionless shaking acceleration $\Gamma \equiv A(2\pi f)^2/g$ remains approximately constant at a low value of $\Gamma \sim 0.01$. Although the mean and the size of the fluctuations in the friction coefficient both remain relatively constant with increasing $f$, the nature of the slips does change significantly with $f$. In particular, as $f$ is increased, the pulling force is increasingly characterized by steady, elastic-like increases and large avalanche-like stress drops. Additionally, the self-affine roughness of the force-versus-distance curves evolves with frequency, with the Hurst exponent~\cite{krim1993prl,krim1993pre} increasing from approximately 0.5 to 0.7 as $f$ is increased from 0 to 30~Hz. To interpret this result, we suggest that increasing frequency enhances stick-slip via increased number of oscillations during shear. In this scenario, the grains become better organized and more commensurate with the rough slider during rearrangements, converting many smaller rearrangements into fewer larger ones. These conditions, particularly the grain mobility induced by the mechanical vibrations, can increase static friction levels~\cite{He1999,Muser2001,Coffey2005} and alter stick slip behavior.

\section{Methods}
\label{sec:1}

Our experimental apparatus, depicted schematically in Fig. 1(a), has been described in detail in an earlier publication~\cite{krim2011stick}. It consists of a solid slider that is pulled at a constant speed over a two-dimensional granular material. The granular bed is roughly 1.5 meters long and 0.15 meters high. The grains are bidisperse photoelastic disks with diameters 4 and 5 mm. The slider has mass $M=0.17$~kg. The bottom edge of the slider is patterned by half-round cutouts of diameter comparable, but not equal to, the diameters of the grains that it is in contact with. The slider is coupled to a translational stage that moves at constant velocity of $v=5$~mm/s via a spring with spring constant $k=160$~N/m. A digital force sensor is used to record the spring force $F$ as a function of distance traveled, which is then converted to a dimensionless friction coefficient $\mu=F/Mg$. Fig. 1(b) shows a typical plot of $\mu$ versus distance.

\begin{figure}
(a) \\  
  \includegraphics[width=\textwidth]{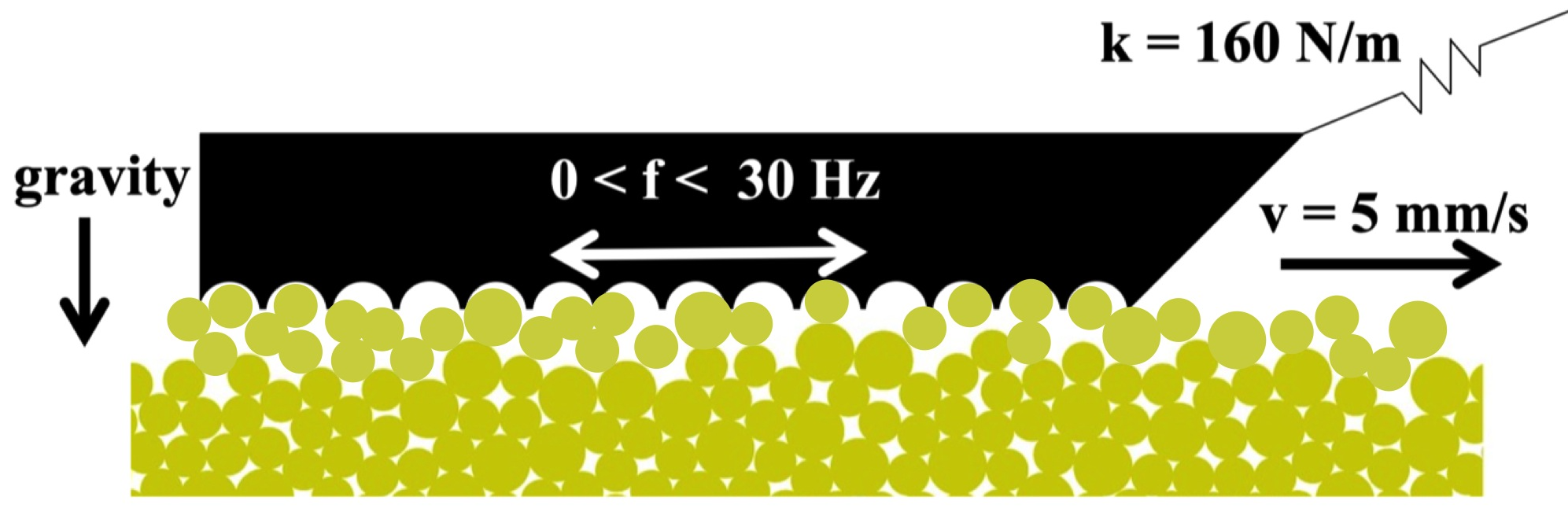} \\
(b) \\    \includegraphics[width=\textwidth]{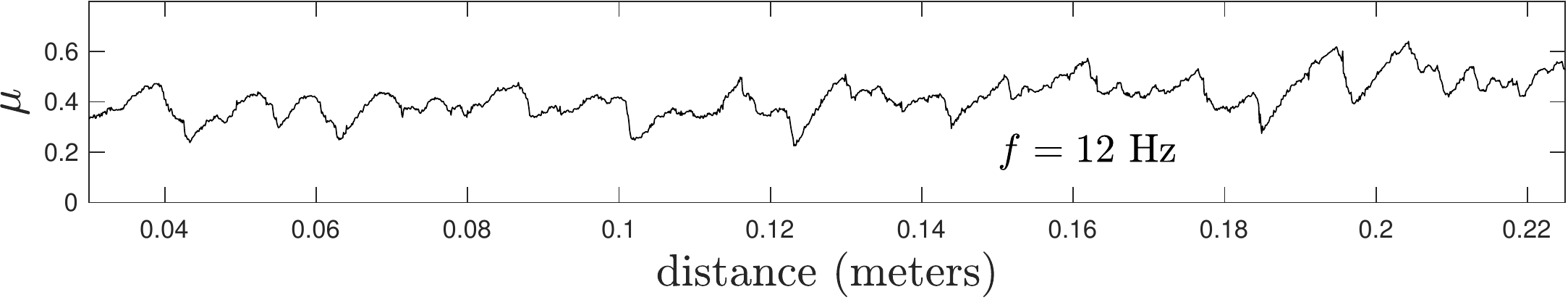} \\
\caption{(a) A schematic of the apparatus. (b) Dimensionless pulling force $\mu \equiv F/mg$ versus distance for one experiment, where the slider is pulled across a granular bed at 5~mm/s and vibrated at 12~Hz. The data in (b) has been notch-filtered to suppress contributions from the 12~Hz signal itself. }
\label{fig:cartoon}       
\end{figure}

\begin{figure}
(a) \\  
  \includegraphics[width=0.32\textwidth]{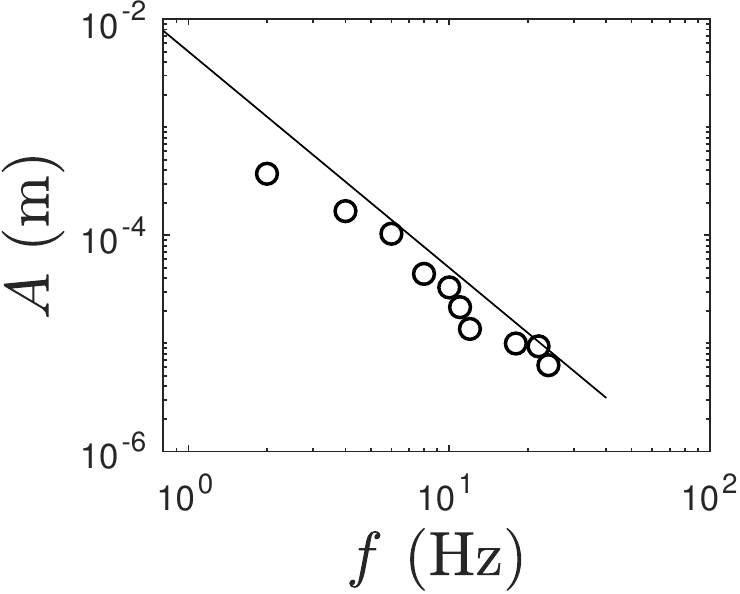}
  \includegraphics[width=0.32\textwidth]{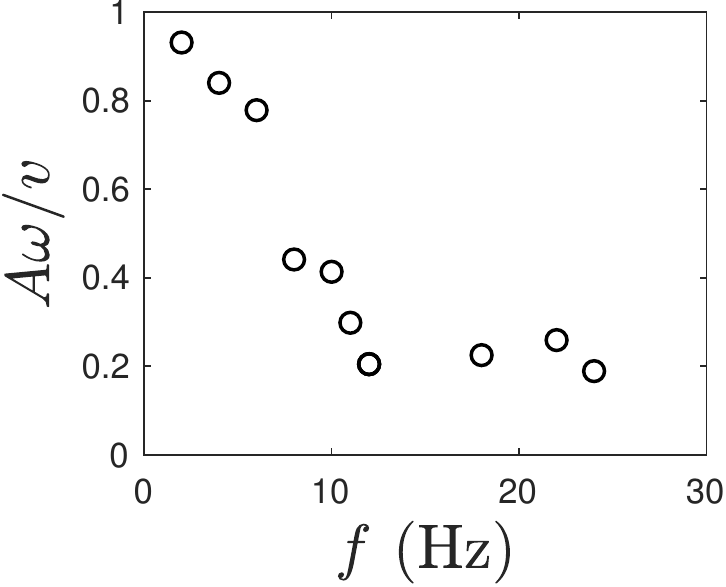}
   \includegraphics[width=0.32\textwidth]{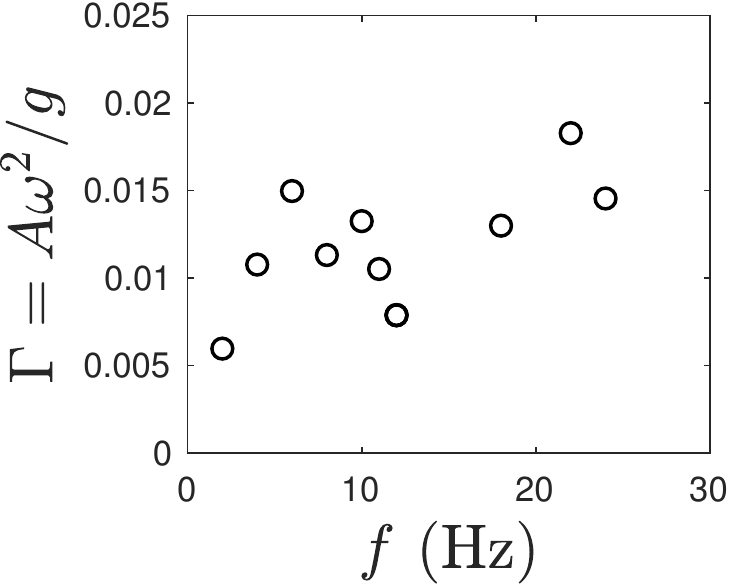} \\
\caption{(a) The amplitude of the applied vibration $A$ plotted versus frequency $f$, with a solid line showing $A \propto f^{-2}$. (b) The dimensionless shaking velocity $A\omega / v$, where $\omega = 2\pi f$ and $v=5$~mm/s is the average speed of the slider, is plotted versus $f$. (c) The dimensionless acceleration $\Gamma \equiv A \omega^2 / g$, where $g$ is the gravitational acceleration, is plotted versus $f$.}
\label{fig:freq-spectrum}       
\end{figure}

Magnets (not shown) are attached to the top of the slider and are positioned within two coils. We then drive oscillating current through the coils using a function generator and an amplifier. This results in oscillatory electromagnetic forces on the magnets, which causes the slider to vibrate in a direction parallel to its motion. We drive the shaker at vibration frequency $f$ and fixed amplitude of the time-varying voltage sent to the shaker. This results in an amplitude $A$ of vibration that decreases as $A \propto f^{-2}$, as shown in Fig.~\ref{fig:freq-spectrum}. We estimate $A$ by taking the Fourier transform of the force data and examining the value at the vibration frequency. Since the spring force is linearly proportional to displacement, we assume the value $A$ in the force signal is linearly proportional to the displacement amplitude. For the data shown here, we vary $f$ from 0 to 30~Hz, and we estimate the dimensionless shaking amplitude to be  $\Gamma = A (2\pi f)^2/g \approx 0.01$. 

The nature of frictional fluctuations is known to be dependent on system parameters, particularly sliding speed~\cite{Luan2004}. For example, Zadeh, et al.~\cite{zadeh2018crackling} employed a very similar apparatus and reported that, in the absence of mechanical vibrations, the system's behavior could be categorized into three different regimes: stick-slip, irregular, and periodic. At very low speeds (e.g., $v\approx 0.1$~mm/s), stick-slip behavior was reported, and at high speeds (e.g., $v\approx 100$~mm/s), inertial or periodic oscillations were observed at the natural frequency of the slider-spring system. At intermediate speeds, (e.g., $v\approx 15$~mm/s), they observed irregular behavior: not purely stick-slip but not yet dominated by periodic inertial oscillations. Our pulling speed of $v=5$~mm/s puts us near the transition between stick-slip and irregular regimes, consistent with prior reports by Krim, et al.~\cite{krim2011stick} employing the same apparatus. 

\section{Results}
\label{sec:2}

\begin{figure}[h!!]
(a) \\ 
  \includegraphics[width=\textwidth]{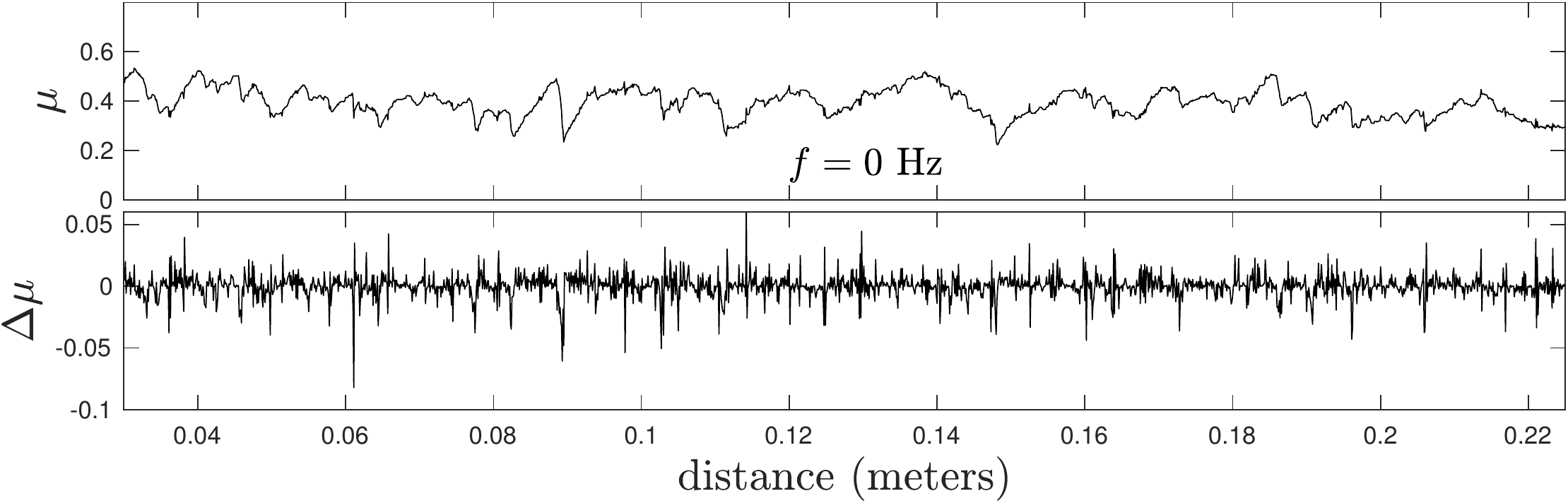}\\
  (b) \\
  \includegraphics[width=\textwidth]{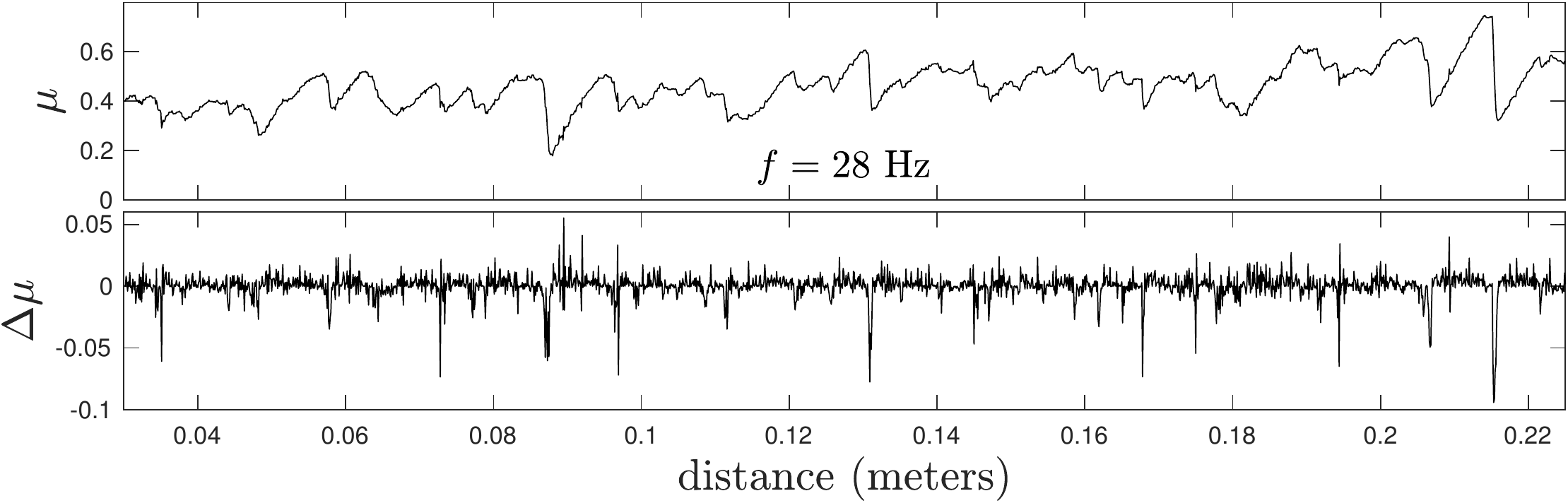}
\caption{Plots of $\mu$ and $\Delta \mu$ (defined as the difference in $\mu$ between each successive pair of data points) as a function of distance for slider velocity of 5~mm/s. Panel (a) shows data with vibration at 0~Hz, and panel (b) shows data with vibration at 28~Hz.}
\label{fig:time-series}       
\end{figure}

\begin{figure}
 \hspace{3mm} (a)  \hspace{33mm} (b) \hspace{33mm} (c) \\ 
 \includegraphics[width=0.33\textwidth]{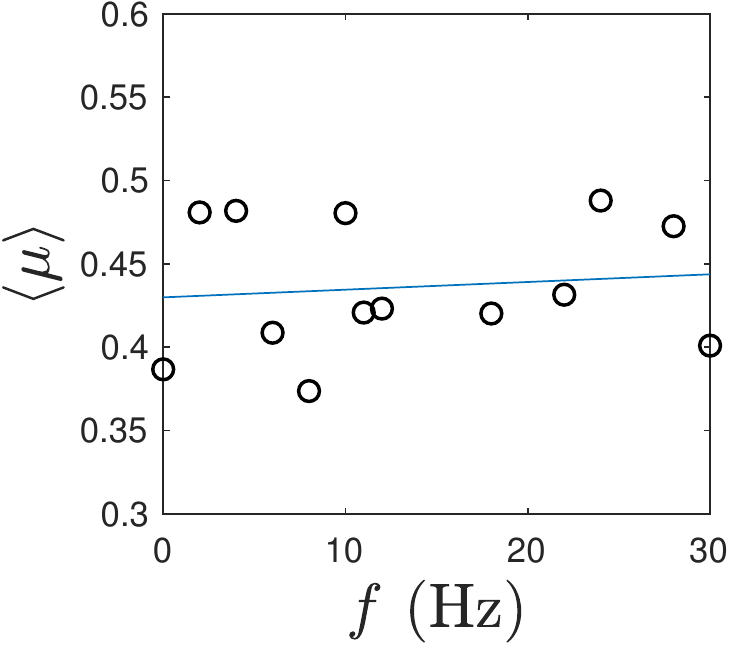}
  \includegraphics[width=0.32\textwidth]{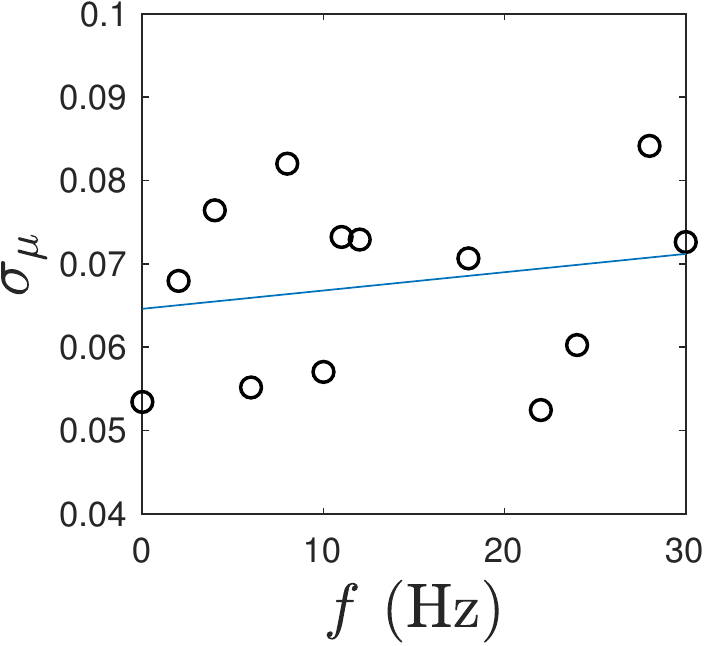}
  \includegraphics[width=0.32\textwidth]{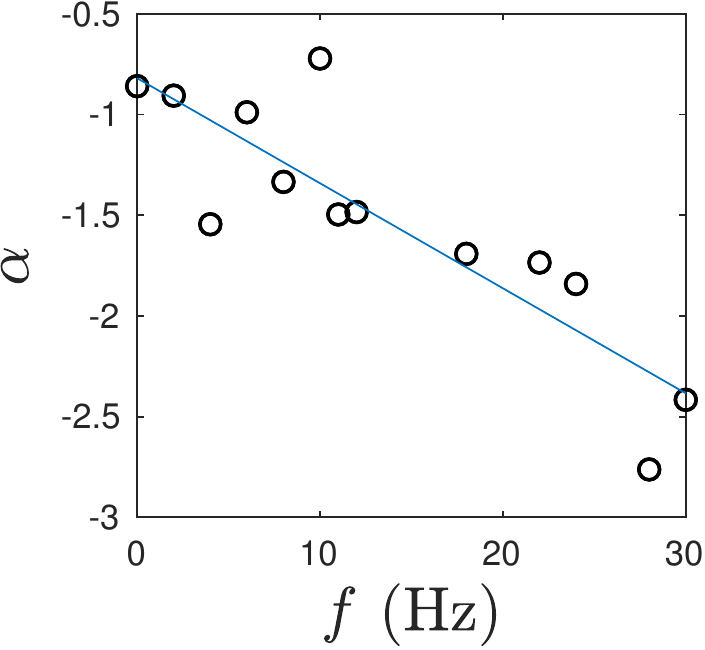}
\caption{(a) The mean dimensionless pulling force $\langle \mu \rangle$, (b) the standard deviation $\sigma_\mu$ of $\mu$, and (c) the skewness $\alpha$ of $\Delta \mu$ are each plotted as a function of vibration frequency $f$. Solid blue lines are linear fits to the data, which show that $\langle \mu \rangle$ and $\sigma_\mu$ are both nearly independent of $f$, but $\alpha$ decreases strongly with $f$.}
\label{fig:stats}       
\end{figure}

In all experiments, $\mu$ begins at zero as the slider is still and the spring is not stretched. As we begin to pull the slider, $\mu$ rises during an initial transient phase, which persists over a short distance that is always less than 3~cm, then reaches a ``steady-state'' phase, where $\mu$ fluctuates around a constant value. We ignore the initial transient phase and focus solely on the steady-state phase. Figure~\ref{fig:time-series} displays typical data segments during the steady-state phase for $\mu$ and $\Delta \mu$, the discrete difference of $\mu$, for no vibration and $f=28$~Hz. We obtain $\Delta \mu$ by subtracting neighboring pairs of data points in $\mu$, such that positive values of $\Delta\mu$ represent increasing $\mu$ and negative values of $\Delta \mu$ represent decreasing $\mu$ (stress drops). Figure~\ref{fig:stats} shows a statistical characterization of the data shown in Fig.~\ref{fig:time-series} for all frequencies studied. As can be seen in Fig.~\ref{fig:stats}(a) the mean pulling force $\langle \mu \rangle$ is virtually independent of $f$: increasing frequency clearly does not reduce the average granular friction for the regime studied here. Additionally, Fig.~\ref{fig:stats}(b) shows that the standard deviation of the friction coefficient, $\sigma_\mu$, is also relatively constant for increasing $f$. 

Despite the fact that the mean and the size of the fluctuations in $\mu$ both remain relatively constant with increasing $f$, the nature of the slips does change significantly with $f$. This can be seen in Fig.~\ref{fig:time-series}, where the data in Fig.~\ref{fig:time-series}(c,d) recorded at the higher frequency of 28 Hz are more asymmetric, with intermittent and larger stress drops. Such features are far less prominent for the data recorded at 0 Hz (Fig.~\ref{fig:time-series}(a,b)). To quantify this asymmetry, we examine the skewness $\alpha$ of the $\Delta \mu$ data sets. This quantity is zero for symmetric dynamics (where stress rises and drops in the same way) and negative for stick-slip dynamics.  Figure~\ref{fig:stats}(c) shows that $\alpha$ is increasingly negative as $f$ increases, ranging from $\alpha \approx -0.8$ at 0 Hz to $\alpha \approx -2.5$ at 30 Hz, reflecting an increasingly asymmetric distribution of $\Delta\mu$ with large, intermittent stress drops. 

\section{Discussion and analysis}
Our results suggest a physical picture where, as vibration frequency increases, many small rearrangements are substituted for a fewer, larger avalanche-like slips. In this scenario, the grains become better organized into more stable configurations, more compacted, and potentially more commensurate with the rough slider during rearrangements. Each of these conditions can increase static friction levels~\cite{He1999,Muser2001,Coffey2005}, alter stick slip behaviors, and increase the skewness of the distribution of $\Delta \mu$. This picture is supported by measurements from Ref.~\cite{krim2011stick} of the same slider being pulled over a compact, solid lattice (i.e., a fixed row of grains), where there are no particle rearrangements and stick-slip dynamics are dominant. In particular, we analyzed data in Ref.~\cite{krim2011stick} from experiments where the slider is pulled over the solid lattice at $v=5$~mm/s, and we found the skewness of $\Delta \mu$ to be -4.5, and increasing to $\alpha \approx 0$ as $v$ is increased. Additionally, $\alpha$ was independent of whether or not we applied a vibration at $f=11$~Hz. We note that for the present study, as frequency is increased, the values of $\alpha$ shown in Fig.~\ref{fig:stats}(c) for sliding over a granular bed are approaching the value associated with a compact, organized substrate. Moreover, the frequency dependence of $\alpha$ is unique to the case of the granular bed, corroborating our assertion that it is associated with grain rearrangements and not surface friction between grains.

Under the physical picture that we suggest, the self-affine nature of the roughness profile~\cite{krim1993prl,krim1993pre} of the force curves would also be highly frequency dependent, as mobile grains would experience an increasing number of vibrations as the slider passed over them, readily contributing to rearrangement.

Self-affine scaling of surface roughness is characterized by
\begin{equation}
\sigma_\mu(L) \propto L^H,
\label{eqn:self-affine}
\end{equation}
where $\sigma_\mu (L)$ is the standard deviation over a sample size $L$, and $H$ is the Hurst exponent, which varies between 0 and 1 and describes the manner in which the roughness scales with lateral extent.
It can morevoer be directly mapped to power spectral analyses common in the literature and force power spectra $P_f (\omega)$ analyses common in the literature~\cite{zadeh2018crackling,zadeh2018seismicity,zadeh2017}, since self-affine force power spectral curves scale as
\begin{equation}
P_f (\omega)\propto \omega^{-(2H+1)}.
\end{equation}
 
\begin{figure}
 \hspace{3mm} (a)  \hspace{33mm} (b) \hspace{33mm} (c) \\ 
  \includegraphics[width=0.33\textwidth]{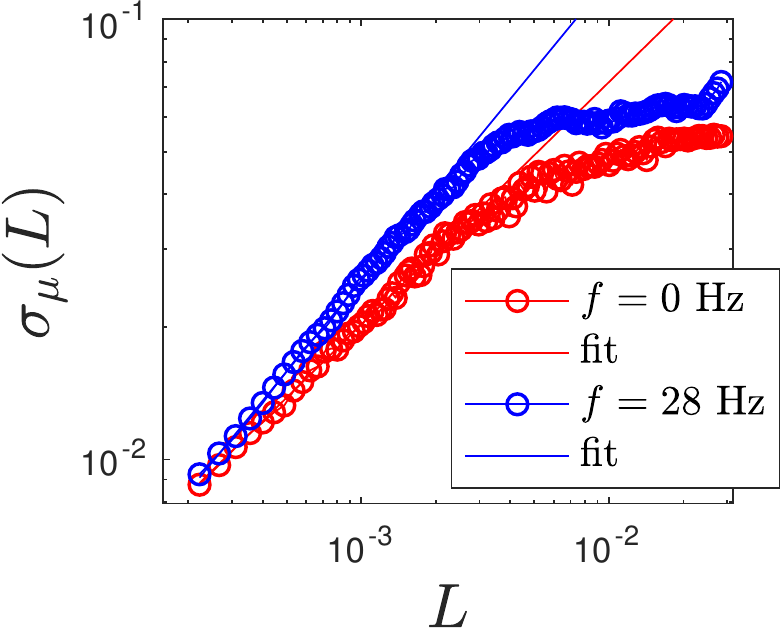}
  \includegraphics[width=0.32\textwidth]{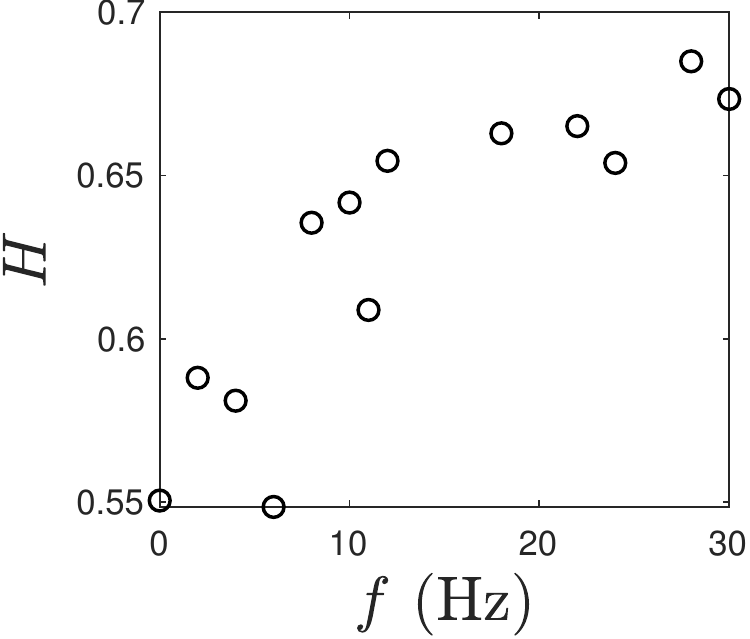}
  \includegraphics[width=0.32\textwidth]{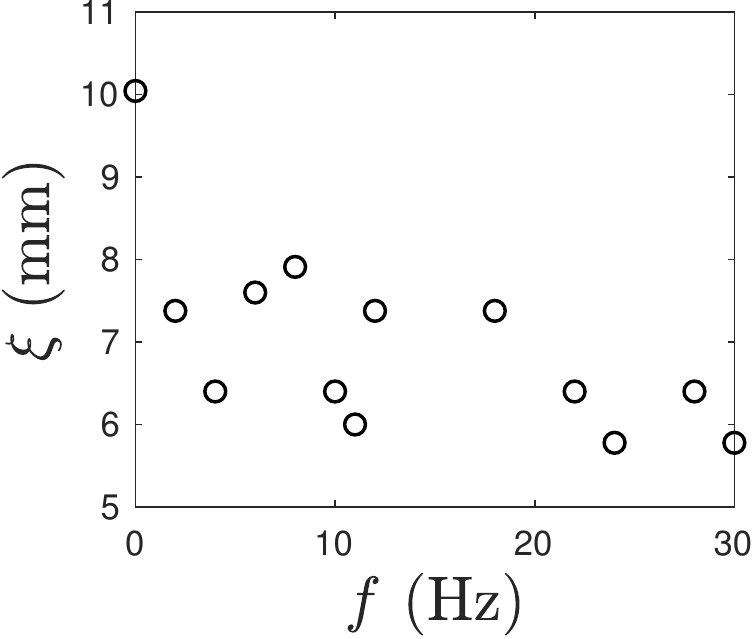}
\caption{(a) The standard deviation $\sigma_\mu$ of $\mu$ is plotted as a function of sample length $L$ for 0~Hz and 28~Hz. The solid lines show fits of the form $\sigma_\mu\propto L^H$ to the small-$L$ portion of the data. (b) The self-affine exponent $H$ increases with vibration frequency $f$. (c) The correlation length $\xi$, defined as the value of $L$ at which the solid fit line shown in (a) is 1.5 times bigger than the data, is plotted versus $f$.}
\label{fig:self-affine}       
\end{figure}

 Figure~\ref{fig:self-affine}(a) shows the dependence of  $\sigma_\mu (L)$ on L for data recorded at 0 and 28 Hz. Both data sets are described by Eq.~(\ref{eqn:self-affine}) for small $L$ and then plateau to a constant value at large $L$, characteristic of self-affine systems. The crossover happens at a correlation length $\xi$, which determines the length scale over which the structures tend to repeat. Increasing $f$ is clearly associated with increases in $H$ (Fig.5(b)), which increases from approximately 0.55 at 0 Hz to 0.67 at 30 Hz. This result is consistent with Fig.~\ref{fig:time-series}, where the high-$f$ data for $\mu$ appears ``smoother'' than the low-$f$ data, and is also close to the H values or 0.5-0.7 inferred from the power spectra data reported by Zadeh {\it et al} \cite{zadeh2018crackling,zadeh2018seismicity,zadeh2017} for unvibrated granular systems. Figure~\ref{fig:self-affine}(c) also shows that the correlation length weakly decreases with $f$, and its value is roughly the same as a particle diameter.

\section{Conclusions}

We have presented here a study of the frictional behavior of a two-dimensional slider pulled slowly over a 
granular substrate comprised of photoelastic disks while being vibrated at frequencies ranging from 
0 to 30 Hz in a direction parallel to sliding. Measurements were performed at speeds close to the transition from stick-slip to steady sliding, where the impact of vibration is likely to be magnified on account of the relative ease which which force chains can be disrupted when close to transitions.  We observe, surprisingly, that increased frequency of vibration can enhance stick-slip behavior for granular friction. Reports of reductions in friction when systems are mechanically vibrated are commonplace in the literature. The apparent paradox is resolved by taking note of the fact that vibrations can induce rearrangements in granular materials that may not otherwise occur in rigid solid systems.

In particular, the results can be explained by a physical picture where, as vibration frequency increases, many small rearrangements are substituted for a fewer, larger avalanche-like slips. At the low acceleration conditions under which the measurements were performed, the  grains become better organized, more compacted and potentially more commensurate with the rough slider during rearrangements. These conditions, and particularly the grain mobility induced by the mechanical vibrations can increase static friction levels~\cite{He1999,Muser2001,Coffey2005} and alter stick slip behaviors. The self-affine nature of the roughness profile~\cite{krim1993prl,krim1993pre} of the force curves would be highly frequency dependent in this scenario, as mobile grains would experience an increasing number of vibrations as the slider passed over them, readily contributing to rearrangement. The force curve profiles are consistent with this interpretation.

\section{Acknowledgements}
This work was supported by NSF DMR0906908 and NSF DMR0805204. 

\bibliographystyle{spphys}       
\raggedright
\bibliography{references.bib}   


\end{document}